\documentclass[twocolumn,prb,epsfig,aps]{revtex4-1}

\usepackage{graphicx}
\usepackage{color}

\newcommand{\beq}{\begin{eqnarray}}
\newcommand{\eeq}{\end{eqnarray}}

\begin{document}
\setcounter{page}{1}

\title{$^3$He on preplated graphite
}

\author{M.C. Gordillo}
\affiliation{Departamento de Sistemas F\'{\i}sicos, Qu\'{\i}micos y Naturales,
Universidad Pablo de Olavide. E-41013 Seville, Spain}

\author{J. Boronat}
\affiliation{Departament de F\'{\i}sica, 
Universitat Polit\`ecnica de Catalunya, 
Campus Nord B4-B5, E-08034 Barcelona, Spain}

\begin{abstract}
By using the diffusion Monte Carlo 
method, we obtained the full phase diagram of $^3$He on top of
graphite preplated with a solid layer of $^4$He. All the $^4$He atoms of
the substrate were explicitly considered and allowed to move during the
simulation.  We found that the ground state is a liquid of
density 0.007 $\pm$ 0.001 \AA$^{-2}$, in good agreement with 
available experimental data.
This is significantly different from
the case of $^3$He on clean graphite, in which both theory and experiment agree
on the existence of
a gas-liquid transition at low densities. Upon an increase in
$^3$He density, we predict a first-order phase transition between a dense
liquid   and a registered 7/12 phase, the 4/7 phase being found
metastable in
our calculations. At larger second-layer densities, a final transition is
produced to an incommensurate triangular phase.

\end{abstract}


\maketitle
\section{Introduction}

Recent heat-capacity studies of $^3$He on
graphite,~\cite{fukuyama1,fukuyama,smart} both clean and preplated,  have reopened
the interest on the phase diagram of this quasi-two dimensional (quasi-2D)
system. In essence, those results confirm a previous hint about the
existence of a stable $^3$He liquid  
at very low densities,~\cite{gasparini}  and are the 
continuation of a wealth of experimental work on this quantum 
fluid.~\cite{greywall,greywall2,godfrin,godfrin2,saunders,godfrin3,
godfrin4,godfrin5,fukuyama} On clean graphite, the introduction of
corrugation in the theoretical description of the surface~\cite{bonninhe3,yo4} 
proved to be the necessary ingredient to produce a
good match  between those calculations and the experimental results of Sato
{\textit{et al.},~\cite{fukuyama} that indicated the existence of a gas-liquid
transition at very low densities. When flat substrates were  considered, no
liquid was found to be stable in most calculations,~\cite{novaco,miller,hnc,
chester,grau,bonin}, but not in all of them.~\cite{brami}

When $^3$He is adsorbed on $^4$He-preplated graphite,
the experimental data of Ref.~\onlinecite{fukuyama} shows a linear
dependence of the heat capacity versus the  second layer density in the
range $\rho \sim 0$-$0.007$ \AA$^{-2}$. This suggests that the ground state of
that $^3$He monolayer is a homogeneous 
liquid with a density fixed by the upper
bound of the interval.   For a smaller number of atoms on the same surface,
the system is fragmented in drops of 0.007 \AA$^{-2}$ 
with enough empty space
between them to produce the average 2D helium concentration we consider.
This is different from what happens in the  
first layer adsorbed directly on graphite~\cite{fukuyama,bonninhe3,yo4} 
in which there is a gas of $\sim  0.006$
\AA$^{-2}$ in equilibrium with  a liquid of $\sim 0.014$ \AA$^{-2}$. 
In this second case, 
what we have are drops of density 0.014 \AA$^{-2}$ surrounded by a gas of
0.006 \AA$^{-2}$ in the right proportions to produce any density in that
range.  When $\rho$ is above 0.007 
and 0.014 \AA$^{-2}$ for the
second and first layers, respectively, we have homogeneous liquids covering
all the surface.   

In this work, we calculate the complete phase diagram of $^3$He on
$^4$He preplated graphite, including the high density solid region. Given
the importance of considering corrugation   in the description of the bare
substrate, we considered explicitly all the $^4$He atoms on the first layer,
allowing them to move during the simulations. This means that  
there
can be deviations between the actual positions of those atoms and their
corresponding crystallographical sites. That produced 
larger error  bars in the
statistical sampling than in the case
of a flat surface, but it was found to be necessary to reproduce the
experimental results, both in the low and  high density regimes.   

The rest of the paper is organized as follows. In Sec. II, we discuss
the quantum Monte Carlo method used for the simulations and account for the
interaction models and trial wave functions used as importance sampling.
The obtained results and their comparison with experimental data are
reported in Sec. III. Finally, Sec. IV comprises the main conclusions of
the present work.

\section{Method}

We rely on a microscopic approach to the physical problem of a second
layer of $^3$He atoms adsorbed on $^4$He-preplated graphite. The
Hamiltonian is written as,  
\begin{eqnarray} \label{hamiltonian}
\lefteqn{H = -\frac{\hbar^2}{2m_4} \sum_{i=1}^{N_4}  \nabla_i^2    -\frac{\hbar^2}{2m_3}
\sum_{i=1}^{N_3}   \nabla_i^2 } \nonumber \\ 		
& &  + \sum_{i=1}^N V_{\text{ext}}({\bf r}_i)   
+ \sum_{i<j}^N V (r_{ij}) \ ,
\end{eqnarray}
where $N_3$ and $N_4$ are the number of $^3$He and  $^4$He atoms
($N=N_3+N_4$), of masses $m_3$ and $m_4$, respectively.
$V_{\text{ext}}({\bf r})$ is the total interaction of each helium atom 
at position {${\bf r}$ 
with all the carbon atoms in the graphite substrate. 
This term is made up of a sum of individual C-He interactions, 
each of them  modeled by the anisotropic Carlos-Cole potential~\cite{carlosandcole,jltp},
even though no influence of the C-He anisotropy is expected in the behaviour of 
the second layer, as has been already proved for a second layer of $^4$He on the same
substrate \cite{jltp}.   
As in previous works, the substrate structure was modeled
by an stack of honeycomb 2D-lattices separated by $3.35$ \AA$ $ in the typical 
A-B-A-B graphite disposition.~\cite{yo,yo2,yo3} $V (r)$ is the
standard Aziz potential for the helium-helium interaction~\cite{aziz} 
of two atoms separated a distance $r$.

We considered unpolarized
$^3$He, i.e., $N_{\uparrow} = N_{\downarrow} = N_3/2$.     
To solve the $N$-body Schr\"odinger equation corresponding to the Hamiltonian of Eq.
\ref{hamiltonian},  we used the fixed-node Diffusion Monte Carlo (FN-DMC)
method.
This stochastic technique solves the Schr\"odinger equation in
imaginary time, providing the exact energy for the ground state of a Bose system and an upper
bound for a Fermi one.~\cite{hammond} This means that our results are only strictly valid 
at 0 K. Our system consists of a mixture of
bosons ($^4$He) and fermions ($^3$He) and thus we get an upper bound to the
exact energy whose quality depends on the proximity of the nodal surface of
the trial wave function (used for importance sampling) to the (unknown)
exact one. Our trial wave function was

\begin{eqnarray}
\Phi({\bf r}_1, {\bf r}_2, \ldots, {\bf r}_N) = \Psi_3({\bf r}_1, {\bf
r}_2, \ldots, {\bf r}_{N_3}) \times \nonumber  \\
\Psi_4({\bf r}_{N_3 + 1}, {\bf r}_{N_3+2}, \ldots, {\bf r}_{N})
\label{trialtot}
\end{eqnarray}
with ${\bf r}_1, {\bf r}_2, \ldots, {\bf r}_{N_3}$ 
the coordinates of the $N_3$ $^3$He atoms in the second layer, and 
${\bf r}_{N_3 + 1}, {\bf r}_{N_3+2}, \ldots, {\bf r}_{N}$} the ones for 
the $N_4$ remaining 
$^4$He atoms in the first layer, in direct contact with graphite. 
This means that, by construction, the two helium layers are separated from each other, what 
we think a reasonable approximation in light of the available experimental results. 
Following Ref. \onlinecite{yo4}, we defined    
\begin{eqnarray}
\Psi_3({\bf r}_1, {\bf r}_2, \ldots, {\bf r}_{N_3}) =  
D^{\uparrow} D^{\downarrow}  \prod_i^{N_3}  u_3({\bf r}_i)  \times \nonumber \\
\prod_{i<j}^{N_3} \exp \left[-\frac{1}{2} 
\left(\frac{b_3}{r_{ij}} \right)^5 \right], 
\label{trial3}
\end{eqnarray}
where  $D^{\uparrow}$ and $D^{\downarrow}$ are the  Slater
determinants of the two-dimensional system defined by our simulation cell,
that depend on the coordinates of the spin-up  and spin-down $^3$He
fermions. The one-body function $u_3({\bf r})$ is the numerical 
solution of the  
Schr\"odinger equation that describes a single $^3$He atom on top of a
triangular lattice formed by  $^4$He atoms located in the crystallographic
positions ($x_{\rm site}$,$y_{\rm site}$) corresponding to an
incommensurate triangular phase.  To solve this equation, we used the same
technique as in Ref.~\onlinecite{yo4}, and neglected the influence of the
underlying  graphite structure.
In the present
work, we used two triangular $^4$He lattices  at different
densities: $0.112$ \AA$^{-2}$ (corresponding to a
separation
of $3.2$ \AA), and $0.12$ \AA$^{-2}$ (with a He-He
lattice constant of $3.1$ \AA). The
first density corresponds to the $^4$He promotion to the second 
layer,~\cite{doble,godfrin4} while the latter 
is the upper experimental  limit
given  in Ref.~\onlinecite{fukuyama}, a limit slightly 
larger than the one
given in Ref.~\onlinecite{godfrin4} ($\rho \sim  0.116$ \AA$^{-2}$). The
parameter $b_3$ in Eq. \ref{trial3} is variationally optimized; its value
is set to   
$b_3
= 2.96$ \AA, as in Refs.~\onlinecite{grau,yo4}.  The coordinates in the
Slater determinants were corrected by backflow terms in the standard way,
\begin{eqnarray}
\tilde x_i  & = & x_i + \lambda \sum_{j \ne i} \exp [-(r_{ij} -
r_b)^2/\omega^2] (x_i - x_j) \\
\tilde y_i  & = & y_i + \lambda \sum_{j \ne i} \exp [-(r_{ij} - r_b)^2/\omega^2] (y_i - y_j).
\end{eqnarray} 

The optimal values for the parameters in the backflow term 
were 
those of the bulk three-dimensional
system,~\cite{casulleras} i.e., $\lambda = 0.35$; $\omega =
1.38$ \AA,  and $r_b = 1.89$ \AA, since they were proved to give lower
energies~\cite{yo4} that those corresponding to a pure 2D system.~\cite{grau}   

The part of the  trial wave function corresponding to the
first $^4$He layer is taken as,
\begin{eqnarray}
\Psi_4({\bf r}_{N_3 + 1}, {\bf r}_{N_3+2}, \ldots, {\bf r}_{N})  = 
\prod_i^{N_4}  u_4({\bf r}_i) \times  \nonumber \\ 
\prod_{i<j}^{N_4} \exp \left[-\frac{1}{2}
\left(\frac{b_4}{r_{ij}} \right)^5 \right]   \times \nonumber  \\
\prod_i^{N_4} \exp \left\{ -a_4 [(x_i-x_{\rm site})^2 + (y_i-y_{\rm site})^2]
\right\}  
\end{eqnarray}
The last (Nosanov) term in the equation pines the atoms around 
their crystallographic
positions. The free parameters in $\Psi_4$ are: 
$b_4 = 3.07$ \AA$ $(as in Refs.~\onlinecite{yo,doble}), and
$a_4$ has the same value that in a previous study of the second layer
of $^4$He on $^4$He    
(0.55 \AA$^{-2}$).~ \cite{doble} As before, $u_4({\bf r}_i)$ is the
numerical solution of the one-body  Schr\"odinger equation for one $^4$He
atom on top of graphite.  

Eq. \ref{trialtot} defines adequately the $^3$He layer when it is a
liquid or a gas. 
If this is not so, we have
to introduce a Nosanov term
to confine the $^3$He atoms,  
\begin{equation}
\prod_i \exp \left\{ -a_3 [ (x_i-x_{\rm site})^2 +  (y_i-y_{\rm
site})^2] \right\} \ ,
\label{trialsol}
\end{equation} 
in which we used the same $a_3$ parameter for all the solid phases and
densities ($a_3 = 0.24$ \AA$^{-2}$).~\cite{yo4} 
We checked that neither an increase or a decrease of up to 50\% in $a_3$ varies
the stability ranges of the solids described.   
The trial wave function (\ref{trialtot}) 
is factorized in a term depending only on the  $^3$He
coordinates and other only on the $^4$He ones. 
Simulations including a Jastrow term relating atoms in both layers produced 
no significant change in the obtained results.

\section{Results} 

In Fig. \ref{liquids} we show the FN-DMC results 
for the energy per
$^3$He atom as a function of the second-layer density.   The upper
(lower) curve
corresponds to a fixed first-layer density of 0.112 (0.12)
\AA$^{-2}$.
We chose those densities as lower and upper limits to the 
magnitude  at hand, and used these two set of simulations to test the effect of a
possible compression of the $^4$He solid structure upon an increase of the number
of $^3$He atoms on top of it. The curves displayed are  third-order degree
polynomials obtained by least-squares fitting to the
shown data.

From those fits, we can say state that the energy per $^3$He atom on
$^4$He-preplated graphite  in the infinite dilution limit, $E_0$, is 
$-24.45 \pm 0.04$ K in the first case, and $-24.74 \pm 0.07$ K for the
denser preplating, values  much smaller in absolute value than
the same magnitude for $^3$He on bare graphite, $-135.771 \pm 0.001$ K.~\cite{yo4} In
both cases, we used a simulation cell that comprised $14 \times 8$
incommensurate triangular lattice cells. Those rectangular unit cells
contain two $^4$He atoms, and have a surface $d_{{\rm He}} \times \sqrt 3 
d_{{\rm He}}$, $d_{{\rm He}}$ being the distance between two atoms in the
first layer. This means that for the upper curve the simulation cell was 
$44.8  \times 44.34$ \AA$^2$, and for the lower one, $43.4 \times 42.95$
\AA$^2$. In both cases, that means 224 $^4$He atoms on the lower layer and 
enough $^3$He atoms to produce the densities shown. This means up to 102
$^3$He
atoms in the case of the second-layer liquids discussed below. 
In all the simulations, standard finite-size corrections to the fermionic wavefunction
were applied. As indicated in the preceding Section, 
all the $^4$He  atoms on the first layer
were allowed to move during the simulations, i.e., we solve the full
Hamiltonian (\ref{hamiltonian}).

Our results, reported in  Fig. \ref{liquids}, show  
second layer is a liquid, since in both cases we have self-bound structures
of densities $0.007 \pm 0.001$ \AA$^{-2}$ (upper curve) and $0.017 \pm
0.001$ \AA$^{-2}$ (lower one). Those values were obtained
from the fits
displayed. 
At difference with the first layer on 
bare graphite structure, no gas-liquid transition 
was found.
To our knowledge, the only experimental set of available
data to
compare our results to are those of Ref. \onlinecite{fukuyama}. As
indicated above, at very low densities, the experimental heat capacity
depends linearly on  the second layer density in the range 0-0.007
\AA$^{-2}$. This would indicate a phase separation between a clean surface
and an homogeneous liquid of 0.007 \AA$^{-2}$. This is exactly what can be 
deduced from the upper curve of Fig. \ref{liquids}. From that, we can
conclude
that a model with a preplated density of 0.112 \AA$^{-2}$ is a good
description of the experimental setup.         

\begin{figure}
\begin{center}
\includegraphics[width=0.8\linewidth]{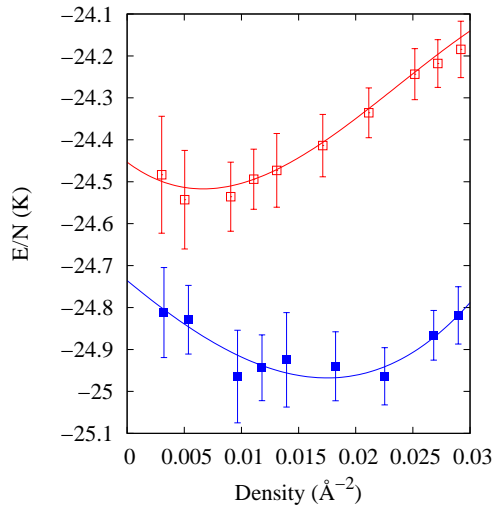} 
\caption{(Color online) Energy per $^3$He atom ($E/N$) on the second layer
of  $^3$He on graphite as a function of the second-layer density. Upper
curve, underlying $^4$He density 0.112 \AA$^{-2}$;  lower curve, first
layer density 0.12 \AA$^{-2}$). The curves are third-order polynomial fits
to the data displayed.        
}

\label{liquids}
\end{center}
\end{figure}

Our results are  
different from the results of a previous
theoretical calculation on this same system.~\cite{bonninhe3}  In that
work, the authors found a gas-liquid transition of the same type that the
one for the first layer on bare graphite. 
To explain that discrepancy,  
we use the results displayed in   Fig.
\ref{comparacion}. There, the lower set of data is the one in the previous
figure after having subtracted the energy in the infinity dilution limit,
$E_0$.  On the other hand, the open circles correspond to a calculation in
which the effect of the first $^4$He layer and the graphite surface below
it have been described by a laterally  $z$-averaged potential, something
similar to the description made in Ref.~\onlinecite{bonninhe3}.  In this
flat case, $E_0 = -26.313 \pm 0.001$ K. Even tough the results in
Ref.~\onlinecite{bonninhe3}  are not exactly the same as the ones in Fig.
\ref{comparacion} since the helium-helium potentials  are slightly
different, both sets of data look pretty similar to each other. At the same time, 
both of them are noticeably 
different from the   
simulations made with the full Hamiltonian (\ref{hamiltonian}). The error 
bars of those results are of the size of the symbols, and not displayed for 
simplicity. 

The missing ingredient is  the corrugation of
the $^4$He substrate, as in the bare graphite case. 
To check that, we repeated our calculations but
without allowing the movement of 
the $^4$He atoms in the first layer. What
we  see is that those results are within the (rather large) error bars of
those corresponding to the full $^4$He-moving calculation, and present a
minimum at approximately the same density.  In our opinion, this indicates
that the introduction  of a $^3$He effective mass,
used in the Ref.~\onlinecite{bonninhe3}  calculations, is not enough to compensate for the use of  a
$z$-averaged potential. The error bars of these last results are comparable, 
though somewhat smaller than the ones for the full calculation, and again were
not displayed for simplicity. 

\begin{figure}
\begin{center}
\includegraphics[width=0.8\linewidth]{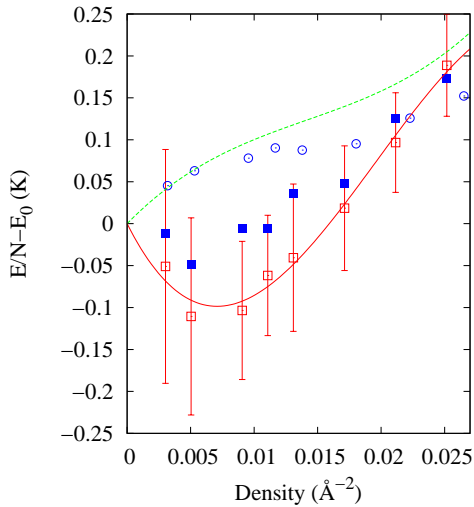} 
\caption{(Color online) 
Same as in the previous figure, but after subtracting the infinite dilution
limit energies ($E_0$'s). Lower open squares with error bars, our
simulation results for $^3$He on top of an
active first $^4$He
layer; upper full squares without error bars, same but for a fixed $^4$He
layer with atoms located in their crystallographic positions; open circles,
results for a second $^3$He layer on a first averaged over potential. The
lower line corresponds to the same third-order polynomial fit of the
previous figure, and the upper one, to the pure 2D result of Ref.
\onlinecite{grau}.      
}
\label{comparacion}
\end{center}
\end{figure}

Fig.~\ref{solido} shows the high density end of the $T=0$ isotherm for a
0.112 \AA$^{-2}$ underlying density. There, we have displayed the energies
per $^3$He atom corresponding to a liquid phase (full squares),
incommensurate triangular solid (open squares), and two standard
commensurate structures on second-layer-helium structures: $4/7$ (open
circle), and $7/12$ (full circle). By a simple inspection, we can see that
there is practically no difference between the energies of the $4/7$
lattice, of density 0.064 \AA$^{-2}$, and that of an incommensurate
triangular structure of the same density. In addition,  the energy per particle of a
$7/12$ structure is lower than both the one corresponding to 
the $4/7$
commensurate solid and that of the incommensurate arrangement of equal
density (0.066 \AA$^{-2}$). This means that we can draw a double-tangent
Maxwell construction line that starts in a liquid of density $0.047 \pm
0.002$ \AA$^{-2}$ and ends in the energy corresponding to that $7/12$
structure,  i.e., those phases are in equilibrium with each other. When the
density increases further, another Maxwell construction 
shows that the $7/12$ solid is in equilibrium with an incommensurate  triangular
one of density $0.072 \pm 0.004$ \AA$^{-2}$, that is probably stable up to
the third layer promotion.

\begin{figure}
\begin{center}
\includegraphics[width=0.8\linewidth]{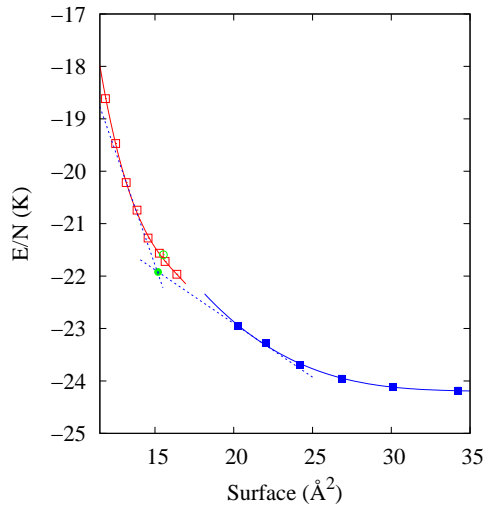} 
\caption{(Color online)
Energy per $^3$He atom ($E/N$) on the second layer of helium on graphite on
top of a 0.112 \AA$^{-2}$ $^4$He layer, as a function of the inverse of the
second layer density.  Solid squares, liquid phase; Open squares,
triangular incommensurate structure; full circle, $7/12$ phase; open
circle, $4/7$ commensurate solid.   Solid lines are mere guides-to-the-eye.
Dotted lines correspond to Maxwell constructions between the different 
stable phases. 
}
\label{solido}
\end{center}
\end{figure}

In the same way as before, to check for the influence of the first layer
density on the phase diagram of the second layer, we performed another
series of calculations for a $^4$He density of 0.12 \AA$^{-2}$. The results
are displayed in Fig.~\ref{solido2}, where the symbols and lines have the
same meaning as in the previous figure. The main differences are the
densities at which  the different transitions are produced: the liquid is
now stable up to $0.054 \pm 0.002$ \AA$^{-2}$, and it is in equilibrium
with a $7/12$ commensurate solid of 0.070 \AA$^{-2}$, that undergoes  a
first-order phase transition to a triangular solid of $\rho = 0.074 \pm
0.002$ \AA$^{-2}$. If, as suggested in Ref.~\onlinecite{godfrin4}, the
density of the first $^4$He layer in a real  setup is between those
considered here, one would expect density limits for the stable phases in
between those reported for the two sets of simulation results.  

With that in mind, we can say that our results compare favorably with the
available experimental data. In particular, a solid is found for second
layer densities larger than~\cite{godfrin3}  0.072 \AA$^{-2}$, a density
which is well
within the error bars of our results in both series of simulation data.
This would correspond to the lower stability limit of the triangular
phase.   In addition, magnetization~\cite{bauphd} and
heat-capacity~\cite{saunders} measurements  
on a double $^3$He layer, give a solid-solid
transition region similar to the one that we see from a commensurate structure
to the lower density for which the incommensurate phase is stable
(total density in the range 0.179 and 0.185 \AA$^{-2}$). 
 However, even though those data are 
close to our results, they are
not directly comparable to them,  since it is well known that  the
density of the first layer of the system $^3$He/$^3$He/graphite
is
$\sim$ 5\% smaller than the $^3$He/$^4$He/graphite one~.\cite{fukuyama3}  
Finally, the experimental upper second layer
density for a liquid in equilibrium with the commensurate phase ($7/12$ in
our case, but $4/7$ in virtually all the previous literature)  
is~\cite{godfrin4} $0.058 \pm 0.005$ \AA$^{-2}$. This is comparable to our
0.12 \AA$^{-2}$ result, what would support a compression of the first layer
upon population of the first one.  

\begin{figure}
\begin{center}
\includegraphics[width=0.8\linewidth]{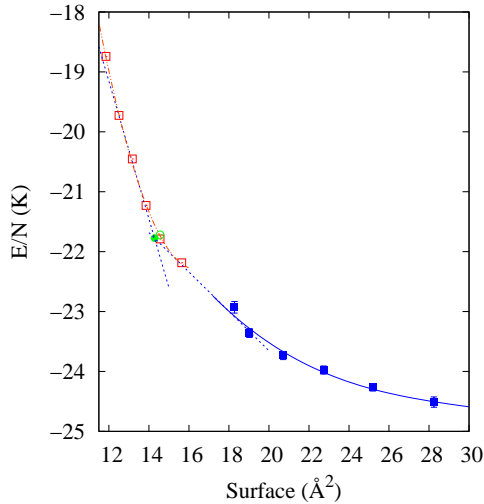} 
\caption{(Color online)
Same as in the previous figure, but for a $^4$He density of 0.12 \AA$^{-2}$. 
}
\label{solido2}
\end{center}
\end{figure}

\section{Discussion} 

In this work, we have calculated the complete phase diagram of a
$^3$He layer 
on top of $^4$He-preplated graphite at 0 K. The main difference with 
previous theoretical approaches
is that we have considered a full
three-dimensional system and allowed the $^4$He atoms of the first layer to move.
This produces a low-density phase diagram that is appreciably different
from
that for a flat~\cite{grau} or $z$-averaged~\cite{bonninhe3} substrate.
This result
is not surprising, since corrugation~\cite{manu1,manu2} and 
disorder~\cite{corboz,doble} 
have been found to be important
in double layer  $^4$He simulations. However, the fact that our results are
comparable to the experimental data of Ref.~\onlinecite{fukuyama} confirms
the quality of our approximations.   The rest of the phase diagram is pretty
close to the experimental data, with phases and
stability ranges that are in good agreement each other. 

However, there is a significant difference. 
The commensurate phase that we found to be
in equilibrium with the dense liquid and the triangular solid is a $7/12$,
not the $4/7$, as it appears in the literature.  To try to understand this,
we should have in mind that the difference in densities between those
two
phases is very small ($\sim$ 3\%, expressed as density of the second layer
only). This means that it could be experimentally difficult to distinguish
them, especially taking into account that the first layer densities can
vary. On the other hand, our Hamiltonian does nor include any ferromagnetic
or antiferromagnetic spin-spin interaction, 
something that could  be
important and eventually change the relative stability of the $4/7$ and $7/12$
phases.     

\acknowledgments
We thank H. Godfrin and H. Fukuyama for illuminating discussions. 
We acknowledge partial financial support from the
Junta de Andaluc\'{\i}a group PAI-205 and and MINECO (Spain) Grants 
No. FIS2014-56257-C2-2-P  and FIS2014-56257-C2-1-P.

\end{document}